\documentstyle[prl,aps,epsf]{revtex}

\tighten
\begin{document}
\draft

\title{Anomalously Localized States in the Anderson Model}

\author{V. M. Apalkov$^1$, M. E. Raikh$^1$, and B. Shapiro$^2$ \\
$^1$Department of Physics, University of Utah, Salt Lake City,
UT 84112, USA \\
$^2$Department of Physics, Technion-Israel Institute of
Technology, Haifa 32000, Israel}
\maketitle

\begin{abstract}
In a diffusive conductor the eigenstates are spread over the entire
sample. However, with certain probability, an anomalously localized
state (ALS) can occur, i.e. the wave function assumes anomalously
large values in some region of space. Existing analytical theories of
ALS are based on models described by a continuous (Gaussian) random
potential.  In the present paper we study ALS in a lattice (Anderson)
model.  We demonstrate that close to the center of the band, $E=0$, a
new type of ALS exist and calculate analytically their likelihood.
These ALS are {\em lattice-specific} and have no
analog in the continuum.  Our findings are relevant to numerical
simulations, which are necessarily performed on a lattice.
We demonstrate that inconsistencies with "continuous" results
reported in the previous numerical work on ALS can be explained within
our analytical theory.
Finally, we point out
that, in order to compare the numerics with the "continuous" ALS
theories, simulations must be carried out not too far from the band
edges, within the band, where the continuous description
applies. Simulations performed for $E$ close to the band center reveal
lattice-specific ALS that do not exist in continuous models.

\end{abstract}

\vspace{5mm}


{\bf \noindent
Introduction.}
In a weakly disordered  conductor the typical value of an
eigenfunction intensity,
$\vert\Psi_{\alpha}\left(\bbox{r}\right)\vert^2$, is of order
$L^{-d}$, where $L$ is the sample size, in $d$ dimensions
($d=2,3$). However, with certain probability, the intensity can
assume anomalously large values. The study of such rare events in
diffusive conductors was
pioneered in Ref. \onlinecite{Kravtsov}
and further pursued in
Refs.\onlinecite{Muzykantskii95,falko95,mirlin00}.
The "prelocalized" states,
studied in
Refs.\onlinecite{Kravtsov,Muzykantskii95,falko95,mirlin00}, exhibit
an anomalous buildup of intensity in some region of space, of a
size larger than the mean free path, $l$. The properties of these
states are universal, in the sense that the disorder enters only
via the mean free path. Another type of rare events was identified and
studied in \cite{apalkov'02}. The corresponding eigenstates,
designated as "almost localized states", are confined primarily to
small rings, of a sub-mean-free-path size. These states are
non-universal, i.e. sensitive to the microscopic details of the
system. In particular, their likelihood sharply increases with the
correlation radius, $R_c$, of the disordered potential (for fixed
value of $l$). In what follows we designate any type of an
anomalously large buildup of intensity as an anomalously localized
state.

The above mentioned analytical studies of the ALS were limited to
models described by a continuous (Gaussian) random potential. On
the other hand,  numerical studies of disordered electronic
systems are necessarily performed on the lattice,
most often within
the Anderson model\cite{anderson58}, with the
tight-binding Hamiltonian
\begin{equation}
\label{Hamiltonian}
\hat{H} = \sum _{ \bbox{r}, \bbox{r}^{\prime }  }
    c^{\dagger } _{\bbox{r}}  c _{\bbox{r}^{\prime }} +
   \sum _{\bbox{r}} V_{\bbox{r}} c^{\dagger } _{\bbox{r}}  c _{\bbox{r}},
\end{equation}
where
$c^{\dagger } _{\bbox{r}}$ is the creation operator
of an electron at site $\bbox{r}$ of a $d$-dimensional
hypercubic lattice with lattice constant equal to 1,
  and
$V_{\bbox{r}}$ is a random on-site energy with r.m.s.
$\Delta_d = \langle V_{\bbox{r}}^2 \rangle^{1/2}$.
The Anderson model has become a powerful tool for numerical study
of various disorder-related phenomena. As the computing
capabilities constantly grow, allowing diagonalization of
large-size matrices, more and more accurate information can be
inferred from the simulations.
As a result, the early
success\cite{mackinnon81} in confirmation of the scaling theory
\cite{abrahams79} was followed by  recent numerical studies
that have successfully addressed more delicate  issues such as
{\em (a)} critical exponents\cite{slevin99}
and  critical behavior of the eigenfunctions  in 3d \cite{grussbach95},
{\em (b)} quantitative characteristics of the quantum Hall
transition\cite{huckestein94},
{\em (c)} different aspects of the
level statistics at the Anderson transition
\cite{zharekeshev97,ndawana02},
{\em (d)} Anderson transition in 2d \cite{asada02} possible
with spin-orbit coupling\cite{hikami80},
{\em (e)} the critical conductance distribution at
the transition \cite{slevin97,markos02},
{\em (f)} verifying scaling for the full conductance distribution
\cite{slevin01}.

This successes have encouraged a number of authors
\cite{uski00,uski01,uski02,nikolic'01,nikolic02,nikolic'02,patra03}
to employ
the Anderson Hamiltonian for
numerical study of the ALS in disordered conductors.
In particular, the subject of interest is the function $f_d(E,t)$
defined as
\begin{equation}
\label{intensities}
f_d(E,t)\! =\!
\frac{1}{\nu L^d}\!
\left\langle \sum_{\alpha}\delta
\left(t \! - \! \vert\Psi_{\alpha}\left(0\right)\vert^2L^{d}\right)
\delta\left(E\! -\! E_{\alpha}\right)\right\rangle,
\end{equation}
where $\Psi_{\alpha}$ and $E_{\alpha}$ are the eigenfunctions and
eigenenergies of the Hamiltonian (\ref{Hamiltonian}), respectively, and
$\nu(E)$ is the  density of states.
ALS are responsible for the large--$t$ tail of the
disorder-averaged distribution
(\ref{intensities})
of the eigenfunction intensity at a given energy, $E$.
They
correspond to the anomalous buildup of certain  eigenfunctions inside
the volume $L^d$
\cite{Kravtsov,Muzykantskii95,falko95,mirlin00}.
Simulations performed have revealed a number of unexpected
peculiarities in the likelihood
of the  almost localized states:
(i) in 2d,
the behavior
$f_2(E,t) \propto \exp\left(-C_E\ln^2t\right)$
which is in accord with theoretical prediction of Refs.
\onlinecite{Kravtsov,Muzykantskii95,falko95,mirlin00}
was 
obtained\cite{uski00}. However, upon changing the disorder
magnitude, $\Delta_2$, the constant $C_E$ did not scale with the
conductance $g \sim \Delta_2^{-2}$;
(ii) the magnetic field dependence of $C_E$ in Ref. \onlinecite{uski00}
turned out to be very weak. This is in  contrast to the
simulations\cite{ossipov02}
of the eigenfuctions intensity distribution,
in which  the model of
the kicked rotator rather than the Anderson model was studied. While
the simulations\cite{ossipov02} have also revealed
$\exp\left(-C\ln^2t\right)$ behavior, the coefficient $C$ in the
presence of the time-reversal symmetry  was almost two times smaller
than in the limit when this symmetry was completely broken;
(iii) Simulations of Ref. \cite{patra03} suggest that
the likelihood
of the almost localized states is non-universal.
Namely,  $f_d(E,t)$ depends on the
{\em correlation radius} of the random on-site energies,
$V_{\bbox{r}}$, for a {\em given} conductance, $g$,
which is, again, in conflict with
theory;
(iv) Simulations in 3d\cite{nikolic'01,nikolic02} indicate
that  the wave function intensities,
$\vert\Psi(m_x,m_y,m_z)\vert^2$, of ALS
in  a cube with a side $L$,
plotted in the lexicographic order
$m \rightarrow L^2(m_x-1) + L(m_y-1) + m_z$,
are {\em structured}; a typical wave function
represents a system of well defined and almost {\em even-spaced}
spikes, each spike extending
over a certain narrow interval of $m$.
In contrast, diffusive wave functions
$\vert\Psi(m)\vert^2$, plotted in the same way,
do not exhibit any structure\cite{nikolic'01,nikolic02}.

In the present paper we demonstrate that the above peculiarities
stem from the fundamental difference between the  ALS
in continuum and on the lattice.
This difference is most pronounced for energies
close to the band center, $E=0$, where all the simulations
\cite{uski00,uski01,uski02,nikolic'01,nikolic02,nikolic'02,patra03}
were performed.

One should realize that the point $E=0$ is rather special and that
"continuous"  theories can break down in the vicinity of this point, even if
the  mean free path is much larger than the lattice spacing. For instance,
in the 1d case, it has been recently explained in Ref.~\onlinecite{deych}
why the single parameter scaling,
well established for weakly disordered continuous models, is violated near
$E=0$.
Consider the Schr\"{o}dinger equation
$\left(E(\bbox{k}) - E \right) A_{\bbox{k}} +
\sum _{\bbox{k}^{\prime}} U_{\bbox{k}-\bbox{k}^{\prime }}
   A _{\bbox{k}^{\prime}} = 0$
for the Fourier components, $A_{\bbox{k}}$, of
wave function, $\Psi $, where
$E(\bbox{k}) =  2\left( \cos k_x
+\cos k_y \right)$ is the bare spectrum in 2d; the Fourier components,
$U_{\bbox{p}}$, of the on-site disorder, $V_{\bbox{r}}$,
are described by the following correlator
$\left\langle U_{\bbox{p}}U_{\bbox{q}}\right\rangle =
\Delta_2 \sum_{\bbox{G}}  \delta _{\bbox{p},\bbox{q}+\bbox{G}} $,
where $\bbox{G}$ are the vectors of the reciprocal lattice.
The terms with $\bbox{G} \neq 0$ are responsible for the
difference between the lattice and continuum. Indeed, as
it was demonstrated in Ref. \onlinecite{deych}, with regard to
these terms, the bare spectrum $E(k)=2\cos k$ in 1d must be
considered as a two-band spectrum $E= \pm 2\cos k $
within the reduced Brillouin zone $\vert k \vert < \pi/2$. In
other words, the umklapp processes, resulting
from the terms $\bbox{G} \neq 0$, are small in parameter $1/l$,
which does not have an analog in continuum.
Away from $E=0$ this parameter is diminished by a factor
$\left(\vert E\vert l\right)^{-1}$ for $\vert E \vert \gg l^{-1}$.
Similar "non-universal" phenomena near $E=0$ should persist also in
lattice models in higher dimension.
In 2d, the center of the band corresponds to the four lines
$k_x\pm k_y =\pm \pi$, which constitute a square (see Fig. 1).
Due to umklapp processes at $\left(0,\pm \pi\right)$
and $\left(\pm \pi, 0\right)$, this square is the ``right''
Brillouin zone, which accommodates the ``upper'' $\left( E > 0 \right)$
and the ``lower'' $\left( E <0 \right)$ bands. The umklapp related
processes are now small in parameter $1/l^2$, which is, again,
specific to the lattice.

Below we identify a
new type of ALS, specific for the lattice, and calculate
analytically their density. The analytical theory allows to account for all
the peculiarities (i)-(iv) uncovered in the numerics.
The main idea is that the intrinsic
periodicity of the Anderson model offers a possibility to localize
the electronic states by the {\em  periodic} fluctuations,
$V_{\bbox{r}}$,  with a period 2, which cause a ``dimerization''
of the underlying lattice. In 1d,
such Peierls-like
fluctuations\cite{peierls} create a gap in the spectrum of the
tight-binding Hamiltonian (\ref{Hamiltonian}).  In order to ``pin''
the center of localization of the in-gap state to a certain
lattice site,
a $\pi$ phase shift in the periodicity should occur at this
site,
by analogy to the
topological solitons\cite{su}. Our prime observation is that
similar fluctuations (period-doubling along each axis accompanied by
a $\pi$ phase shift) are capable to localize tight-binding electron
in higher dimensions, {\em without formation of a gap}.
It is still convenient pedagogically to start from the 1d case,
and then
generalize the theory to higher dimensions.

\vspace{3mm}
{\bf \noindent
1d case.} In the absence of disorder,
the bare density of states is
$\nu(E)=\pi^{-1}\left(4-E^2\right)^{-1/2}$.
As was explained above, the period-doubling fluctuation
$V^{(0)}(n\neq 0) =
V_0 \mbox{sign}\left(n\right)\exp\left(i\pi n\right)$
with
$\pi$ phase shift at $n=0$ creates a localized state with energy,
$E \in \left[-V_0, V_0\right]$.
Indeed, upon introducing a vector
$\mbox{\large$\chi$}_n=\Bigl\{\Psi\left(2n\right), \Psi\left(2n-1\right)\Bigr\}$
the Schr\"{o}dinger equation $\hat{H} \Psi  = E\Psi$ with on-site
energies $V(n)$ takes the form
\begin{equation}
\label{chi1}
\left(E - V_0 \sigma_3  -\sigma _1 \right)\mbox{\large$\chi$}_n =
\sigma _{-}
 \mbox{\large$\chi$} _{n+1} + \sigma _{+} \mbox{\large$\chi$} _{n-1},
\end{equation}
where $\sigma_i$ are the Pauli matrices. The corresponding eigenvector
$\mbox{\large$\chi$} _n  \propto
\exp\left(i \pi n - 2 \gamma _1 \vert n \vert\right)$
decays  both to the left and to the right from
the site $n=0$\cite{su} with a decrement
$\gamma _1  (V_0)=\mbox{arcsinh}\sqrt{\frac{1}{4}
\left(V_0 ^2 - E^2\right)}$.
The actual position of the localized state within the gap
$\left[-V_0, V_0\right]$ is governed by the on-site energy at the
origin, $V(n=0)$.
Namely,  $V(n=0)$ and $E$ are related as
$V(n=0)=E\left[1 + e^{-\gamma _1}/\sinh\gamma _1\right]$.

For a 1d interval of a length, ${\cal L}$, the fluctuation
$V^{(0)}(n)$ with
$V^{(0)}(0)= V(n=0)$
results in the buildup,
$\tau = |\Psi(0)|^2/|\Psi({\cal L}/2)|^2 =
   \exp\left[2\gamma _1(V_0){\cal L}\right]$,
of the eigenfunction at $n=0$. However,
the statistical weight of the fluctuation $V^{(0)}(n)$ is zero.
Random deviations of the on-site energies, $V(n)$,
from $\pm V_0$ give rise to
a certain
distribution, ${\cal P}_{1,{\cal L}}(E,\tau)$, of buildups, $\tau$.
To find this distribution, we note that
deviations of $V^{(0)}(n)$ from $\pm V_0$ result in the
fluctuations of the
{\em local} decrement, $\gamma _1$. Then
the expression
for ${\cal P}_{1,{\cal L}}$ can be written as
\begin{eqnarray}
& & {\cal P}_{1,{\cal L}} (E,\tau) =
\int_0^{\infty } dV_1  \int_{-\infty }^{0} dV_2
\ldots
P(V_1) \ldots P(V_{{\cal L}/2})~\!  \nonumber \\
       & & ~~\times
   \delta \left(\tau -  e^S \right)=
\left\langle \delta \left(\tau - e^S \right) \right\rangle _{V_n}=
 \frac{1}{\tau} \tilde{\cal P}_{1,{\cal L}}
\left(E,\ln \tau \right),
\label{1d_expression}
\end{eqnarray}
where  $S=2\sum_{n}\gamma _1(V_n)=2\sum_n
    \mbox{arcsinh}\sqrt{\frac{1}{4}
\left(V_n ^2 - E^2\right)}$, and  $P(V_i)$ is the distribution function
of energy  of the site $i$.
To proceed further, we notice that the function
$\tilde{\cal P}_{1,{\cal L}}$
satisfies the following recurrent relation
\begin{equation}
\label{recurrent}
\tilde{\cal P}_{1,{\cal L}} (\ln \tau) =
\int d V_1 P(V_{1})~\!
  \tilde{\cal P}_{1,{\cal L}-1}
\Bigl( \ln \tau  - 2 \gamma _1\left(V_{1}\right) ~\!\Bigr).
\end{equation}
It is easy to express the solution of Eq.~(\ref{recurrent}) in
terms of auxiliary function
\begin{equation}
\label{auxiliary}
I_1(\kappa, E ) = \int_E^{\infty }
   dV P(V) \exp \left[ -2 i \kappa \gamma _1 (V)
 \right].
\end{equation}
Then we have
\begin{equation}
\tilde{\cal P}_{1,{\cal L}}(\ln \tau) \! = \!\!
  \int d \kappa ~  e^{i\kappa \ln \tau}~
\left[I_1(\kappa)\right]^{\mbox{\tiny ${\cal L}$/2}}
= e^{-{\cal F} _{1,{\cal L}} (E, \tau)}.
\label{integral}
\end{equation}
The analytical expression for log-probability of the buildup,
${\cal F} _{1,{\cal L}} (E, \tau)$, can be obtained in the
domain of ${\cal L}$ where the
the main contribution to the integral (\ref{auxiliary})
comes from small $\kappa $. Then we can use the expansion
$I_1(\kappa ) = I_1(0)  - i
\left\langle \gamma _1\right\rangle  \kappa -
\left\langle\gamma _1^2\right\rangle \kappa ^2$,
where
$\left\langle \gamma _1 (E) \right\rangle$ and
$\left\langle \gamma _1^2 (E) \right\rangle$ are the average
decrement $\gamma _1$ and $\gamma _1^2$, respectively.
Substituting this expansion into (\ref{integral}), we
readily obtain
\begin{equation}
\label{F_1d}
{\cal F} _{1,{\cal L}} =
\frac{ \ln ^2 (\tau /{\cal T}_1) I_1^2(0) }{{\cal L} \left[
 2 I_1(0) \left\langle \gamma _1^2 \right\rangle -
\left\langle \gamma _1 \right\rangle ^2
    \right]}
 + \frac{{\cal L}}{2} \mbox{\Large$|$}\!
 \ln \left\{ I_1(0,E) \right\} \!\mbox{\Large$|$} ,
\end{equation}
where $\ln {\cal T}_1 = \frac{1}{2}
\left\langle \gamma _1 \right\rangle {\cal L}/I_1(0)$.
The domain of applicability of Eq.~(\ref{F_1d}) is
$\Delta _1 {\cal L} \gtrsim \ln (\tau /{\cal T}_1) \gg \Delta _1
{\cal L}^{1/2}$.
The origin of ${\cal T} _1$ and of the second term in
Eq.~(\ref{F_1d}) is the ``prefactor'' in the functional
integral (\ref{1d_expression}).

\vspace{3mm}
{\bf \noindent
2d case}. Our main finding is that,
in 2d,
with the bare density of states
$\nu(E)=
\pi^{-2} \left( 1+ \frac{\left| E
\right|}{4} \right) ^{-1}
   \mbox{\bf K}\left( \frac{4 -  \left| E \right|}{4+  \left| E \right|}
\right),
$ where $\mbox{\bf K}$ is the
elliptic function of the first~kind,~so~that $\nu (E)
\propto
 \ln |E|^{-1}$ at $E\!\! \rightarrow \! \! 0$,
 a straightforward extension of the 1d approach
applies. Namely, the period-doubling fluctuation $V(n_x,n_y) = V_0
[ \mbox{ sign} (n_x) \exp \left(i \pi n_x\right) +
                     \mbox{sign} (n_y) \exp\left(i \pi n_y\right)]$
creates an ALS
{\em without opening a gap} in $\nu(E)$.
Indeed, with $V(n_x, n_y) = V(n_x) + V(n_y)$ being {\em additive},
the solution of the 2d tight-binding equation is {\em multiplicative},
i.e. $\Psi (n_x, n_y) \propto
\exp\left[ \left(i \pi/2  - \gamma _2\right)  \left(n_x + n_y \right)
\right]$. The  decrement $\gamma _2 (V_0)=
\mbox{arcsinh}\sqrt{\frac{1}{16} \left(4 V_0 ^2 - E^2\right)}$
differs from $\gamma _1$ due to the fact that, in 2d, the energy $E$
is the sum of energies of motion along $x$ and $y$.
Now, analogously to the 1d case, we consider a square with a side,
${\cal L}/\sqrt{2}$, as shown in Fig.~1, and
introduce a distribution, ${\cal P}_{2,{\cal L}} (E,\tau )$, of the
probability that the buildup from the perimeter to the center
along {\em each path} exceeds $\tau $.
The reasoning leading to recurrent relation between
$\tilde{{\cal P}}_{2,{\cal L}}(E, \ln \tau)=
\tau ^{{\cal L}} {\cal P}_{2,{\cal L}} (E,\tau )$
and $\tilde{{\cal P}}_{2,{\cal L}-1}$  goes as follows. The
perimeter site, $i$, of the square, ${\cal L}$, is connected to
perimeter sites of the square, ${\cal L}-1$, with
one horizontal and one vertical link, as illustrated
in Fig.~1. Denote with $\tau _{i}^{({\cal L}-1)}$ and
$\tau _{i-1}^{({\cal L}-1)}$ the values of buildup from these two
perimeter sites to the center. The evolution of $\tau $
along the horizontal and vertical links can be expressed as
$\ln \tau _{h,i}^{(\cal L)} = \ln \tau _{i}^{({\cal L}-1)} + 2
\gamma _2 ^{(h,i)}$, and
$\ln \tau _{v,i}^{(\cal L)} = \ln \tau _{i-1}^{({\cal L}-1)} + 2
\gamma _2 ^{(v,i)}$, respectively.
This leads to the following relation
\begin{eqnarray}
\label{recurrent_2}
& & \tilde{\cal P}_{2,{\cal L}}\left( \left\{ \ln \tau_{h,i},
\ln \tau_{v,i} \right\}
 \right)  =   \int \ldots \int \left\{
dV_{i}P(V_{i})
\right\}   \times  \nonumber \\
& &
~~~
 \tilde{\cal P}_{2,{\cal L}-1}\left( \left\{
\ln \tau_{i} - 2 \gamma_2^{(h,i)},
 \ln \tau_{i-1} - 2 \gamma_2^{(v,i)}
\right\}
 \right).
\end{eqnarray}
To make Eq.~(\ref{recurrent_2}) closed, we recall that the
actual buildup, $\tau _i$ is the {\em minimal} of the horizontal
and vertical values, i.e.
$\tau _i = \min \{\tau _{h,i},\tau _{v,i} \}$.
Taking this fact into account, the solution for
$\tilde{\cal P}_{2,{\cal L}}$ has the form similar to
Eq.~(\ref{integral}) for the 1d case
\begin{equation}
\label{2d_P}
\tilde{\cal P}_{2,{\cal L}}(\ln \tau) \! = \!
\left( \int\!  d\kappa
e^{i \kappa \ln \tau }
\left[ I_2(\kappa )\right]^{\mbox{\tiny ${\cal L}$/2}}
     \right)^{\mbox{\tiny${\cal L}/2$}}\!\!\! \! = \!
e^{-{\cal F}_{2,{\cal L}}(E, \tau )},
\end{equation}
where the function $I_2(\kappa)$ is defined by Eq. (\ref{auxiliary})
with $\gamma_1(V)$ replaced by $\gamma_2(V)$.
Using the small--$\kappa $ expansion
$I_2(\kappa ) = I_2(0)
- i \left\langle \gamma _2\right\rangle  \kappa -
\left\langle\gamma _2^2\right\rangle \kappa ^2$,
 we arrive at the following 2d generalization of Eq.~(\ref{F_1d})
\begin{equation}
\label{generalization}
{\cal F} _{2,{\cal L}} =
\frac{ \ln ^2 (\tau /{\cal T}_2) I_2^2(0) }{ 2 \left[
 2 I_2(0) \left\langle \gamma _2^2 \right\rangle -
\left\langle \gamma _2 \right\rangle ^2
    \right]}
 + \frac{{\cal L}^2}{4} \mbox{\Large$|$}\!
 \ln \left\{ I_2(0,E) \right\} \! \mbox{\Large$|$},
\end{equation}
where $\ln {\cal T}_2 = \frac{1}{2}
\left\langle \gamma _2 \right\rangle {\cal L} \sim \Delta _2 {\cal L}
/I_2(0)$.
The remaining task is to express the intensity distribution
(\ref{intensities})
through the distribution of buildups, $\tau $. We consider
only the 2d case.
For a given  sample size, $L$, and the fluctuation
size, $\cal L$, the values $t$ and $\tau$ are related via
the normalization condition for
$\Psi(n)$, which
can be presented as $tL^{-2} \left[ (2\gamma _2)^{-2} +
\tau ^{-1} (L ^2 - {\cal L}^2 )\right]=1$.
Taking into account that $2 \gamma _2 {\cal L} = \ln \tau $, we obtain
\begin{equation}
\label{t_tau}
 \frac{\tau }{t} =  1- \left(  \frac{{\cal L}}{L} \right)^2 +
\left(\frac{{\cal L}}{L}\right)^2   \frac{\tau }{\ln^2 \tau}
\approx
1 + \left(\frac{{\cal L}}{L}\right)^2   \frac{\tau }{\ln^2 t} .
\end{equation}
Expressing $\tau $ from Eq.~(\ref{t_tau}) and substituting into
Eq.~(\ref{generalization}), we get
\begin{equation}
\label{F_2_t}
{\cal F} _{2,{\cal L}} = \frac{C_E}{4}
\left[ \ln t -
\ln  \left( 1-
\left(\frac{{\cal L} \sqrt{t}}{L \ln t} \right)^2
     \right) - \frac{1}{2} \left\langle \gamma _2 \right\rangle
 {\cal L}
\right]^2
 + \frac{{\cal L}^2}{4} \mbox{\Large$|$}\!
 \ln \left\{ I_2(0,E) \right\} \! \mbox{\Large$|$},
\end{equation}
where
\begin{equation}
\label{C_E}
C_E = \frac{ I_2^2(0) }{ 2 \left[
 2 I_2(0) \left\langle \gamma _2^2 \right\rangle -
\left\langle \gamma _2 \right\rangle ^2
    \right]}.
\end{equation}
Further steps depend on the relation between $L$ and $t$.
For $L \gg \sqrt{t}/\left\langle \gamma _2 \right\rangle \sim
\sqrt{tg} $ the second logarithm in Eq.~(\ref{F_2_t}) containing
 $\frac{{\cal L} \sqrt{t}}{L \ln t}$
can be neglected. Then we get
$ f _{2} (E, t) = 4 \min _{{\cal L}} {\cal F} _{2,{\cal L}}$,
where the factor 4 accounts for the four quadrants.
Performing minimization, we obtain $f_2(E,t) = \eta C_E \ln ^2 t$,
where $\eta < 1$ is a numerical factor. For the Gaussian
distribution $P(V)$ this factor is equal to $\eta = \left[
1 + \frac{1}{4 (\pi -2 ) \ln 2 } \right]^{-1}\approx 0.76  $.
Optimal value of ${\cal L}$ satisfies the condition ${\cal L} \ll
L \ln t /\sqrt{t}$, so that the above assumption is justified.
This assumption implies that the normalization of $\Psi (n)$
is determined by the region $L- {\cal L}$, i.e. outside the
fluctuation.
In the opposite case, $L \ll \sqrt{tg}$, both terms in Eq.~(\ref{t_tau})
have the same order. In this case, the smallest possible
fluctuation size ${\cal L} \sim L \ln t /\sqrt{t}$ should be
substituted into Eq.~(\ref{F_2_t}). This yields
\begin{equation}
\label{f_2_t}
f _{2} = \left[ C_E
\left( 1 -  \frac{ \left\langle \gamma _2 \right\rangle L}{2 \sqrt{t}}
 \right)^2   +  \left( \frac{ 2 L}{\sqrt{t}} \right)^2
 \mbox{\Large$|$}\!
 \ln \left\{ I_2(0,E) \right\} \! \mbox{\Large$|$}
\right] \ln^2 t .
\end{equation}
Since $C_E \sim g $ and
$\left\langle \gamma _2 \right\rangle \sim g^{-1/2}$, it is easy
to see that both $L$-dependent terms in Eq.~(\ref{f_2_t}) are small
compared to $1$. Thus, we again arrive at $f_2  = C_E \ln ^2 t$
with $C_E$ given by Eq.~(\ref{C_E}).
%
%
%

We now return to the peculiarities in the numerical
results listed in the Introduction, and discuss them in light
of the picture of ALS based on the period-doubling fluctuations.
(i) The  dependence of $C_E$ on the disorder strength,
calculated from Eq. (\ref{generalization}) for Gaussian distribution
with a standard notation for the r.m.s.,
$\Delta _2 = \left\langle
V^2 \right\rangle ^{1/2} = 12^{-1/2} W$, is shown in Fig.~2(a).
 For the range of
$2< W< 4$, studied in Ref.\onlinecite{uski00}, the slope of $\ln C_E$
versus $\ln W^{-2} \sim \ln g$ varies within the range $0.67$--$0.83$
for the energy interval $\vert E \vert < 2$. As seen in the inset,
the scaling
$C_E \propto g $ is recovered from (\ref{generalization})
at $W \lesssim 0.5$.
(ii) Insensitivity of $C_E$ to the magnetic field: for the
period-doubling fluctuation  along both axes, $V(n_x, n_y)$, considered
above, this insensitivity is an
immediate consequence of the fact, that the magnetic phases in
hopping matrix elements can be formally  ``absorbed'' into the
phase factors in the on-site values of the
eigenfunction $\Psi (n_x,n_y)$.
The underlying reason for such gauging out is that the fluctuation
$V(n_x,n_y)$ is {\em separable}.
(iii) Sensitivity of $f_d(E,t)$ to the correlation
of the disorder for a given conductance: it is obvious qualitatively
that for the correlation radius, $R_c$, exceeding the lattice constant,
the likelihood of the period-doubling fluctuation,
with $\sim {\cal L}^2$
sign changes of on-site energies at neighboring sites, is drastically
suppressed. On the quantitative level, depletion of the large--$t$
tail in $f_2(E,t)$, observed numerically in
Ref. \onlinecite{patra03} can be
estimated as $\left[ \exp \left( -\frac{\gamma_2^2}{W^2}\right)
\right]^{{\cal L}^2 R_c^2}
\sim \left[f_2(E,t)\right]^{R_c^2}$, so that for $R_c \gg 1$ the
effect is indeed dramatic. The meaning of the power $({\cal L}R_c)^2$
is that maintaining the period-doubling order requires for each site
to ``pay the price'' to all its $\sim R_c^2$ neighbors.
(iv) In 3d, the corresponding period-doubing
fluctuation has the form $V(n_x,n_y,n_z)=
V(n_x)+ V(n_y) + V(n_z)$, with
$V(n) = V_0\mbox{sign}(n)\exp(i\pi n)$, so that, $|\Psi (n)|^2
\propto \exp \left[ -2 \gamma_3 ( |n_x| + |n_y| + |n_z| )\right]$.
In lexicographic presentation\cite{nikolic'01,nikolic02} this
decay manifests itself as a system of prominent quasi-periodic peaks
with a period close to $2L^2$. In
simulations\cite{nikolic'01,nikolic02} the side of the cube
was small, $L_3= 12$. Then the ALS extends over the entire system.
In Fig.~2(b) we present the lexicographic plot of the
{\em analytical} solution $\vert\Psi(n)\vert^2$ for $L_3 = 12$ and
$\gamma _3 = 0.8$. We find that the
shape of $\vert\Psi(n)\vert^2$
is remarkably close to that in numerics of
Ref. \onlinecite{nikolic'01,nikolic02}.





The main message of the present paper is that, in order to test
numerically, within the Anderson model, the  predictions concerning the ALS
of "continuous" theories, simulations must be carried out  not too far from
the band edges ($E=\pm 4$ in 2d and $E=\pm 6$ in 3d)
where the continuous  description applies.
Simulations performed for $E$  close to the band center reveal
{\em lattice-specific} ALS that do not exist in
continuous models.

We acknowledge the support of the
National Science Foundation under Grant No. DMR-0202790 and of the
Petroleum Research Fund under Grant No. 37890-AC6.


\begin{figure}
\centerline{
\epsfxsize=3.2in
\epsfbox{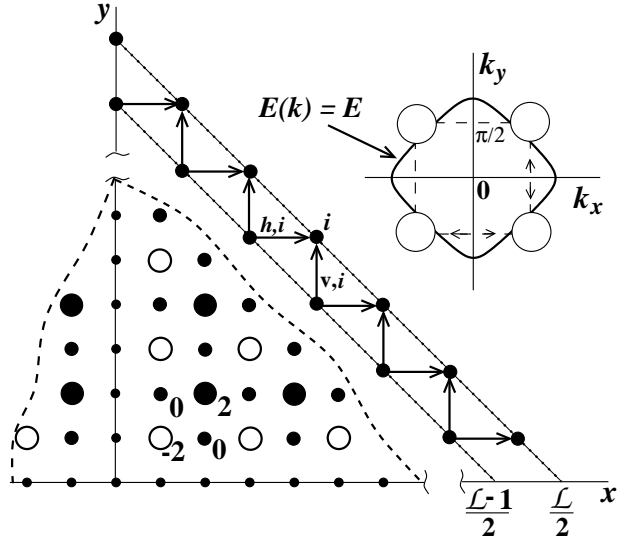}
\protect\vspace*{0.1in}}
\protect\caption[sample]
{\sloppy{ Shown is one quadrant of square with a side
${\cal L}/\sqrt{2}$. On-site energies in the presence of
a period-doubling fluctuation are shown in the units of
$V_0$; $\pi$-phase slip is shown only in the $x$-direction.
Inset: solid line is a surface $E(\bbox{k}) = E$.
Dotted lines are regions in $\bbox{k}$-space perturbed
by the fluctuation.
 }}
\label{fig1}
\end{figure}

\begin{figure}
\centerline{
\epsfxsize=4.0in
\epsfbox{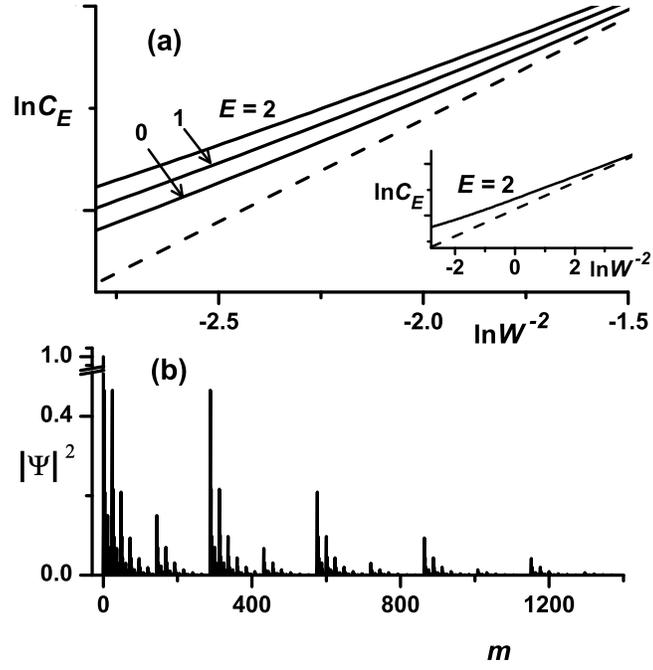}
\protect\vspace*{0.1in}}
\protect\caption[sample]
{\sloppy{ (a) Numerical coefficient $C_E$ in the
dependence $|\ln f_2 (E,t)| = C_E \ln ^2 (t/{\cal T}_2)$
is plotted from Eq.~(\ref{generalization}) versus the disorder
strength, $W$, at different energies. Dotted line
is a weak-disorder asymptotics, $C_E \propto W^{-2}$.
Inset: $C_E(W)$ at $E=2$ is shown in the domain of the
weak disorder. (b) Wave function of an ALS with a decrement
$\gamma _3 = 0.7$ in a cube with a side $L=12$ is
shown in lexicographic order $m \rightarrow
L^2 (m_x-1)+L(m_y-1)+m_z$.
 }}
\label{fig2}
\end{figure}



\begin{references}

\bibitem{Kravtsov} B. L. Altshuler, V. E. Kravtsov, I. V. Lerner,
in {\em Mesoscopic Phenomena in Solids}, eds. B. L. Altshuler,
P. A. Lee, and R. A.~ Webb (North Holland, Amsterdam, 1991).

\bibitem{Muzykantskii95} B. A. Muzykantskii and D. E. Khmelnitskii,
Phys. Rev. B {\bf 51},  5480 (1995).

\bibitem{falko95} V. I. Fal'ko and K. B. Efetov,
Phys. Rev. B {\bf 52}, 17413 (1995).

\bibitem{mirlin00} A. D. Mirlin, Phys. Rep. {\bf 326}, 259 (2000).






\bibitem{apalkov'02} V. G. Karpov, Phys. Rev. B {\bf 48}, 4325 (1993);
V. M. Apalkov, M. E. Raikh, and B. Shapiro,
                             Phys. Rev. Lett. {\bf 89}, 126601 (2002).


\bibitem{anderson58}P. W. Anderson, Phys. Rev. {\bf 109}, 1492 (1958).

\bibitem{mackinnon81}A. MacKinnon and B. Kramer,
Phys. Rev. Lett. {\bf 47}, 1546 (1981).

\bibitem{abrahams79} E. Abrahams, P. W. Anderson, D. C. Licciardello,
and T. V. Ramakrishnan, Phys. Rev. Lett. {\bf 42}, 673 (1979) .

\bibitem{slevin99} K. Slevin and T. Ohtsuki, Phys. Rev. Lett.
  {\bf 82}, 382 (1999).

\bibitem{grussbach95}  H. Grussbach and M. Schreiber,
 Phys. Rev. B {\bf 51}, 663 (1995), and references therein.

\bibitem{huckestein94} B. Huckestein and L. Schweitzer,
Phys. Rev. Lett. {\bf 72}, 713 (1994).


\bibitem{zharekeshev97}I. Kh. Zharekeshev and B. Kramer,
Phys. Rev. Lett. {\bf 79}, 717 (1997).

\bibitem{ndawana02}M. L. Ndawana, R. A. R\"{o}mer,
and M. Schreiber, Eur. Phys. J. B {\bf 27}, 399 (2002), and
references therein.

\bibitem{asada02}Y. Asada,  K. Slevin, and T. Ohtsuki,
Phys. Rev. Lett. {\bf 89}, 256601 (2002), and references therein.

\bibitem{hikami80} S. Hikami, A. I. Larkin, and
Y. Nagaoka, Prog. Theor. Phys. {\bf 63}, 707 (1980).

\bibitem{slevin97}K. Slevin and T. Ohtsuki, Phys. Rev. Lett.
{\bf 78}, 4083 (1997), and references therein.


\bibitem{markos02}P. Marko\v{s}, Phys. Rev. B {\bf 65}, 104207 (2002),
and references therein.

\bibitem{slevin01}K. Slevin, P. Marko\v{s}, and T. Ohtsuki,
Phys. Rev. Lett. {\bf 86}, 3594 (2001); Phys. Rev. B {\bf 67},
155106 (2003).


\bibitem{uski00} V. Uski, B. Mehlig, R. A. R\"{o}mer,
and M. Schreiber, Phys. Rev. B {\bf 62}, R7699 (2000).

\bibitem{uski01}V. Uski, B. Mehlig,
and M. Schreiber, Phys. Rev. B {\bf 63}, 241101(R) (2001).

\bibitem{uski02}V. Uski, B. Mehlig, and M. Schreiber,
Phys. Rev. B {\bf 66}, 233104 (2002).


\bibitem{nikolic'01}B. K. Nikoli\'{c}, Phys. Rev. B {\bf 64},
014203 (2001).



\bibitem{nikolic02}B. K. Nikoli\'{c}, Phys. Rev. B {\bf 65},
012201 (2002).

\bibitem{nikolic'02}B. K. Nikoli\'{c}, V. Z. Cerovski,
Eur. Phys. J. B {\bf 30}, 227 (2002).

\bibitem{patra03} M. Patra, Phys. Rev. E {\bf 67}, 065603(R) (2003).

\bibitem{ossipov02} A. Ossipov, T. Kottos, and T. Geisel,
Phys. Rev. E {\bf 65}, R055209 (2002).



\bibitem{deych} L. I. Deych, M. V. Erementchouk, A. A. Lisyansky,
and B. L. Altshuler, preprint cond-mat/0304440.

\bibitem{peierls}R. E. Peierls, {\em Quantum Theory of Solids}
 (Clarendon, Oxford, 1955), p. 108.

\bibitem{su} W. P. Su, J. R. Schrieffer, and A. J. Heeger,
Phys. Rev. Lett. {\bf 42}, 1698 (1979).






\end{references}
\end{document}